\documentstyle[12pt,aaspp4]{article}
\tighten
\lefthead{Kolb and Tkachev}
\righthead{Femtolensing by Axion Miniclusters}

\begin{document}

\title{Femtolensing and Picolensing by Axion Miniclusters}

\author{Edward W. Kolb}
\affil{NASA/Fermilab Astrophysics Group\\
Fermi National Accelerator Laboratory, Batavia, IL 60510, and \\
Department of Astronomy and Astrophysics, Enrico Fermi Institute\\
The University of Chicago, Chicago, IL 60637}

\author{Igor I. Tkachev}
\affil{Department of Physics, The Ohio State University,
Columbus, OH 43210, and \\
Institute for Nuclear Research of the Academy of Sciences of Russia
\\Moscow 117312, Russia}

\def\be{\begin{equation}}
\def\ee{\end{equation}}
\def\bea{\begin{eqnarray}}
\def\eea{\end{eqnarray}}
\def\la{\mathrel{\mathpalette\fun <}}
\def\ga{\mathrel{\mathpalette\fun >}}
\def\fun#1#2{\lower3.6pt\vbox{\baselineskip0pt\lineskip.9pt
        \ialign{$\mathsurround=0pt#1\hfill##\hfil$\crcr#2\crcr\sim\crcr}}}
\def\rhoeq{{{\rho_{\rm eq}}}}
\def\rhodm{{{\rho_{\rm DM}}}}
\def\Teq{{{T_{\rm eq}}}}
\def\aeq{{{a_{\rm eq}}}}
\def\bra{{{\bar{\rho}_a}}}

\begin{abstract}
Non-linear effects in the evolution of the axion field in the early
Universe may lead to the formation of gravitationally bound clumps of
axions, known as ``miniclusters.''  Minicluster masses and radii
should be in the range $M_{\rm mc}\sim10^{-12} M_\odot$ and $R_{\rm mc} \sim
10^{10}$cm, and in plausible early-Universe scenarios a significant
fraction of the mass density of the Universe may be in the form of
axion miniclusters.  If such axion miniclusters exist, they would have
the physical properties required to be detected by ``femtolensing.''
\end{abstract}



\section{Introduction}

Gravitational lensing is now a well established field of astronomy,
with a wide variety of astrophysical and cosmological applications.
Among the many interesting and important implications of lensing
phenomena is the possibility of interference effects between lensed
images of point-like astrophysical sources
(\cite{m81}; \cite{o83}; \cite{ss85}; \cite{dw86}; \cite{pf91}).  If the
angular separation of the
images of a point-like source is comparable to the wavelength of the
light, interference between the lensed images will lead to a
distinctive fringe pattern both in coordinate space and in energy
spectrum.  The interference pattern in the energy spectrum will depend
upon the location of the detector.  Since well separated detectors
will see different patterns, there is a distinctive signature of the
effect.

If gamma-ray bursters are of cosmological origin they would be small
enough to act as point-like sources, and lenses of extremely small
masses, $10^{-16}\,M_\odot \la M \la 10^{-13}\,
M_\odot$,\footnote{Post-WKB effects may extend the possible detection
range of the lens up to $10^{-11}\, M_\odot$ (\cite{ug95}).} would
produce a frequency-dependent signal in their spectrum
(\cite{g92}; \cite{spg93}).  For source and lens at cosmological distances,
the angular separation of the images would be in the femto-arcsec
range, so the phenomenon was dubbed ``femtolensing'' by Gould (1992).

For source and lens of cosmological distances, the Einstein ring
radius is $R_E \sim  3\times10^{10} (M/10^{-12}M_\odot)^{1/2}$cm.
This means that if two gamma-ray burst detectors are separated by
distances larger than $R_E$, typically one detector will be within the
Einstein ring and the other will not.  Measurement of different fluxes
in two widely separated detectors would extend the detectable range of
lens masses to $M <10^{-7}\, M_\odot$ (\cite{ng95}), which can be called
picolensing.  Thus,
femtolensing and picolensing could be a probe of dark matter in a mass range
difficult
to probe by other means, not to mention settle the issue of the origin
of gamma-ray bursters.

Although the phenomenon of femtolensing (henceforth we omit picolensing
for brevity, but everywhere below we mean both phenomena) is quite intriguing,
the possibility has not attracted a great deal of attention because of a
lack of attractive femtolens candidates. Among the front runners in an
admittedly weak field are snowballs, black holes, and small molecular
clouds.  Snowballs (lumps of primordial H and He) are generally
believed to evaporate on time scales shorter than the Hubble time
(\cite{rjm92}) and for this reason they are not very attractive.  A
significant fraction of closure density in black holes in the
requisite mass range seems difficult to arrange without a very
peculiar spectrum of primordial fluctuations to prevent too many small
holes which would be evaporating now.  As an unattractive possibility
in a field of even more unpalatable choices, small molecular clouds
are usually considered the best bet for femtolensing.

Even if one thought of a way to arrange baryons into femto-mass
objects, it is unlikely that they would contribute a large enough
fraction of closure density for a reasonable optical depth for
femtolensing.  Even if {\it all} baryonic matter would form
femtolenses, their density is constrained by big-bang nucleosynthesis
to be less than about 10\% of closure density.

Since non-baryonic dark matter is expected to dominate the mass
density of the Universe, perhaps it is easier to arrange a near
critical density of {\it non-baryonic} material into objects capable
of femtolensing. But most particle dark matter candidates are very
weakly interacting, and they would not be expected to clump on the
femtolensing mass and density scales.

In this paper we shall show that the situation may be different if
axions are the dark matter.  Here we outline a scenario where a
significant cosmological density of axion miniclusters
(\cite{hr88}; \cite{kt93}, 1994a) form with the necessary properties to be
natural candidates both for femto- and picolensing.

\section{Formation of Axion Miniclusters}

The invisible axion is among the best motivated candidates for cosmic
dark matter. The axion (\cite{sw78}; \cite{fw78}) is the
pseudo-Nambu--Goldstone boson resulting from the spontaneous breaking
of a $U(1)$ global symmetry known as the Peccei--Quinn, or PQ,
symmetry introduced to explain the apparent smallness of strong
CP-violation in QCD (\cite{pq77a}).

There are stringent astrophysical
(\cite{dktw78}; \cite{fwy82a}; \cite{hy88}; \cite{t88}; \cite{rs88};
\cite{m88}) and
cosmological (\cite{pww83}; \cite{as83}; \cite{df83}) constraints on the
properties of
the axion. In particular, the combination of cosmological and
astrophysical considerations restrict the axion decay constant $f_a$
and the axion mass $m_a$ to be in the narrow windows $10^{10}\, {\rm
GeV} \leq f_a \leq 10^{12}\, {\rm GeV}$, and $10^{-5}\, {\rm eV} \leq
m_a \leq 10^{-3}\, {\rm eV}$.\footnote{ For a review see \cite{t90} or
\cite{r90}.}  The contribution to the mean density of the Universe
from axions in this window is guaranteed to be cosmologically
significant.  Thus, if axions exist, they will be dynamically
important in the present evolution of the Universe.

The axion is effectively massless well above the QCD confinement
temperature of a few hundred MeV.  At the confinement scale and below,
QCD effects produce a potential for the axion of the form
$V(\theta)=m_a^2f_a^2[1-\cos(\theta)]$, where the axion field was parametrised
as a
dimensionless angular variable $\theta$.
Above the QCD scale there is no reason for the axion field to be at
its low-temperature potential minimum, so as the axion potential turns
on the field must relax to the minimum of the potential.
Thus the energy density in axions corresponds to
coherent scalar field oscillations, driven by a displacement of the
initial value of the field (the ``misalignment'' angle) away from the
eventual minimum of the temperature-dependent potential.

There will be nothing special in the subsequent evolution of the
axion field if the initial misalignment angle
is constant throughout the Universe.  This may in fact be a reasonable
picture if the Universe
underwent inflation with the ``re-heat'' temperature well below $f_a$.
However, equally plausible early-Universe scenarios lead to
fluctuations of order unity in the misalignment angle on scales larger than the
Hubble radius as the potential starts to develop.  Since the final energy
density depends upon the initial
misalignment angle, spatial fluctuations in the misalignment angle are
transformed into spatial fluctuations in the axion density, which
later lead to tiny gravitationally bound ``miniclusters.''

It is easy to understand that miniclusters will be very dense objects.
Background axion density
scales with temperature as $\bra(T) = 3\Teq s/4$ for $T \ll
\Lambda_{\rm QCD}$, where $s\propto T^3$ is the entropy density and
$\Teq$ is the temperature of equal matter and radiation energy
densities.
Consider a region with axion over-density,
$\rho_a=(1+\Phi) \bra$, where $\Phi$ is larger than unity.
The matter density in that region will
dominate the radiation density at a temperature $T_\Phi =
(1+\Phi)\Teq$.   At this time the density fluctuation became non-linear,
separate out from the cosmological expansion, gravitationally
collapse, relax, and form a minicluster with the approximate density
it had at $T_\Phi$.  A detailed study of this leads to a final
minicluster density of (\cite{kt94b})
\begin{equation}
\label{rhofl}
\rho_{\rm mc} \simeq 140 \Phi^3 (1+\Phi) \bar{\rho}_a(T_{\rm eq}) \approx 3
\times 10^{-14} \Phi^3(1+\Phi)\left(\Omega_ah^2\right)^4{\rm g~cm}^{-3}.
\end{equation}
Even a relatively small increase in $\Phi$ is important because the final
density depends upon $\Phi^4$ for $\Phi\ga 1$.

In Hogan \& Rees (1988) it was assumed that typical values of $\Phi$
corresponding to a minicluster are of order unity.  The typical mass of
a minicluster in Hogan \& Rees (1988) was assumed to correspond to the mass of
all axions inside the horizon at $T \sim 100$ MeV, with the
result $M_{\rm mc} \sim 10^{-5}\, M_\odot$.

In our numerical investigations of the dynamics of the axion field
around the QCD epoch (\cite{kt93}, 1994a) we found that as the
oscillations commence, important, previously neglected, non-linear
effects of the field self-interaction can result in the formation
of transient soliton-like objects
we called axitons.  The non-linear effects result in regions with
$\Phi$ much larger than unity, possibly as large as several hundred,
leading to enormous minicluster densities.  We also found that the
minicluster mass scale is set by the total mass in axions within the
Hubble radius at a temperature around $T =T_1 \approx 1$ GeV when
axion mass is equal to $H$.  This lowers the minicluster mass
from the estimate of Hogan and Rees to about $10^{-12}M_\odot$.

Our previous calculations suggested that miniclusters could be a
candidate for femtolensing.  We have since extended and improved our
original calculation of the properties of axion miniclusters using
the same approach as in \cite{kt93}, but with better
numerical resolution and using a different spectrum of initial
fluctuations of $\theta$. The new calculation confirmed the
basic picture of minicluster formation and sharpened the predictions
of the masses, radii, and abundance of miniclusters. In
the next section we summarize the results relevant for femtolensing.

\section{Physical Properties of Axion Miniclusters}

The evolution of the axion field in an expanding, spatially flat
Universe during the epoch when axion mass switches
on  is governed by a wave-like equation with a non-linear, time-dependent
potential (\cite{kt93}):
\begin{equation}
\psi^{\prime\prime} - \Delta^2\psi + \eta^{n+3}\sin (\psi /\eta )=0 \,\, ,
\label{eq3}
\end{equation}
where $\psi \equiv \eta\theta$, $\psi^{\prime\prime} \equiv d^2\psi/d\eta^2$,
$\eta$ is the conformal time  (normalized to $\eta=1$ when the inverse
of the axion mass is equal to the Hubble radius---i.e.,
$\eta(T_1)\equiv1$), and ${\Delta}$ is the Laplacian
with respect to comoving coordinates.
The factor of $n$ in this equation arises from the way the potential
develops due to instanton effects: $m_a^2(\eta) =
m_a^2(\eta=1)\eta^n$, with $n=7.4\pm 0.2$ (\cite{gpy81}).

As an initial conditions in the new calculation we considered the possibility
that the main source of axions is provided by the decay of axion
strings prior to $T = T_1$ (\cite{d86}; \cite{hs87}; \cite{bs94}).  This might
be
expected if the reheating temperature after inflation is larger than
the temperature of the Peccei-Quinn phase transition.  In such a
scenario the spectral energy density of axions produced by string
decay corresponds to a power law: $d\rho_a/d\omega \propto
\omega^{-1}$, where $\omega=|k|$ (recall that the axion effectively is
massless prior to $\eta = 1$). This corresponds to $|a(k)|= {\rm
const} \times k^{-5/2}$ for the Fourier amplitudes of the axion field.
In our latest calculations we used initial conditions at $\eta = 1$
with amplitudes of the Fourier modes corresponding to the above
spectrum and with random phases. Overall amplitude was chosen to ensure rms
value
of $\theta$ to be $\pi/\sqrt{3}$.

The final density distribution of axions in one of the 2-dimensional slices
through the integration volume is shown in Figure 1.  Three high amplide
density
peaks which are seen on this figure results from non-linear interaction and are
absent in the case of the harmonic potential. They form miniclusters of
interest
at $T \sim \Phi T_{\rm eq}$. The evolution of
the axion field with the above initial conditions was qualitatively
similar to the evolution using the white-noise initial conditions of
our previous calculations.  From the results of the numerical
calculation we determined the mass, radius, and abundance of axion
miniclusters.

\begin{figure}
\plotone{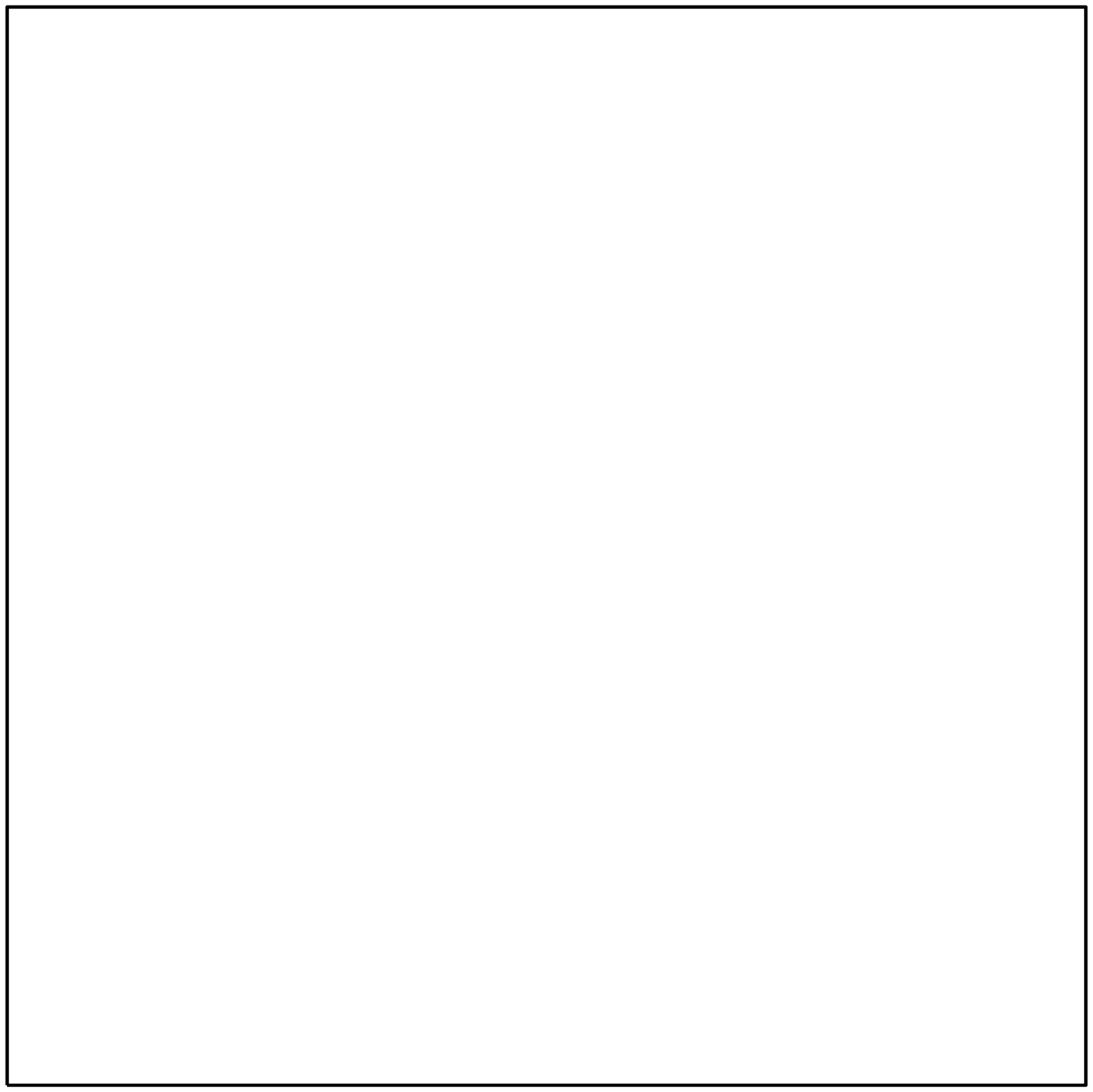}
\caption{This is a two-dimensional slice through a three dimensional
box showing the distribution of axion energy densities for $T \gg T_{\rm eq}$.
The height of
the plot corresponds to $\Phi=\delta \rho_a/\bar{\rho}_a =20$, and the
width to a length of $4H^{-1}(T_1)$ (this corresponds to a
comoving length of about 0.2 pc).  \label{fig1}}
\end{figure}

\placefigure{fig1}

{\it 3.1. Minicluster mass. }
The mass of axion minicluster is set by the scale when axion
oscillations commence. As mentioned above, we have defined $T_1$ as a
temperature when the axion mass equals the Hubble constant, $m_a(T_1)
= H(T_1)$. Using results of Turner (1986) we use $T_1 = (f_a/10^{12} {\rm
GeV})^{-0.175}\, (\Lambda_{\rm QCD}/ 200{\rm MeV})^{0.7} $ GeV. In
what follows we assume $T_1 = 1$ GeV. As a reference mass scale we
define the mass of all axions in a cubic volume of length equal to the
Hubble length at $T_1$, and density equal to the mean cosmological
density: $M_1 = \pi^2 g_*(T_1) T_1^3 T_{\rm eq}/ 30 H_1^3$, where
$g_*(T_1) \approx 62$ is the number of an effectively massless degrees
of freedom at temperature $T_1$. Using $H =1.66 g_*^{1/2} T^2/M_{\rm
Pl}$ we find $M_1 = 8.2 \times 10^{-12} \Omega_a h^2 \, M_\odot$.

Our numerical integration of equation (\ref{eq3}) shows that there is a
spectrum of minicluster masses, but typically the mass of the axion
minicluster is a reasonable fraction of $M_1$, $M_{\rm mc} \la 0.1
M_\odot$.  We estimate the average minicluster mass to be of order
$M_{\rm mc} \sim 10^{-12} \, M_\odot $.

Masses of miniclusters are relatively insensitive to the particular
value of $\Phi$ associated with the minicluster.

{\it 3.2. Minicluster radius. }
Typical axion miniclusters have a large density contrast, $\Phi \ga 1$
(\cite{kt93}), see Figure 1.  Equation (\ref{rhofl}) gives the
minicluster radius as a function of $M$ and $\Phi$:
\begin{equation}
R_{\rm mc} \approx 2 \times 10^{11} \Phi^{-1} \left(1+\Phi\right)^{-1/3}
\left(\Omega_a h^2\right)^{-1} \left(\frac{M}{10^{-12} M_\odot}\right)
^{1/3} {\rm cm} \, .
\label{R}
\end{equation}

The radius of a particular minicluster depends upon its associated
value of $\Phi$.  In the limit $\Phi\ga 1$ the radius of a minicluster
scales as $\Phi^{-4/3}$.

{\it 3.3. Minicluster abundance.}
{}From the numerical calculation we have extracted the mass fraction of
axions in density peaks of $\Phi$ larger than a given value
$\Phi_0$. It is plotted for several values of conformal time $\eta =
1, 3,$ and $4$ in Figure 2.  By the time $\eta>4$ we are no longer able
reliably to resolve the highest density peaks because density contrasts
became too large for our $256^3$ grid. Although at $\eta = 4$ the
evolution has
not completely frozen out, (especially at the large-$\Phi$ end) we can
set a lower bound for the density fractions, e.g., more than 13\% are
in miniclusters with $\Phi > 10$ and 70\% of all axionic dark matter
are in miniclusters $(\Phi > 1)$, see Figure 2.

\placefigure{fig2}

\section{Miniclusters as Femtolens}

For axion miniclusters (or, for that matter, any other possible
candidate) to be detected by femtolensing (or picolensing) the following three
conditions must be satisfied:

{\it i.)} The minicluster has to be in the mass range, $10^{-16}
\,M_\odot \la M_{\rm mc} \la 10^{-11}\, M_\odot$ (\cite{g92}; \cite{ug95})
for femtolensing, or  $M_{\rm mc} \la 10^{-7}\, M_\odot$ (\cite{ng95})
for picolensing.

{\it ii.)} The physical radius of the minicluster has to be smaller
than its Einstein ring radius.  For source and lens
of cosmological distance the Einstein ring radius is, see e.g.
Press \& Gunn (1973):
\begin{equation}
R_{\rm E} \la 3 \times 10^{10} h^{-1/2}
\left( \frac{M}{10^{-12}M_\odot}\right)^{1/2}{\rm  cm},
\label{re}
\end{equation}
where as usual $H_0=100h$ km s$^{-1}$Mpc$^{-1}$.

{\it iii.)} The third condition is that a significant fraction of the
total mass density of the Universe must be in the form of
miniclusters, say $\Omega_{\rm mc} \sim 0.1$, for the lensing rate to be
reasonable. This condition is independent upon the lens mass (\cite{pg73}).

These conditions are non-trivial, which is why there were no serious
candidates for femtolensing.  But we shall now demonstrate that all
three conditions are naturally met for the case of axion miniclusters.

That miniclusters satisfy the mass condition follows from the fact
that the minicluster mass is determined by the total mass of axions
within the Hubble radius at $T\sim1$ GeV.  Of course there will be
some spread in minicluster masses, but as discussed in the previous
section, the masses should be peaked around $10^{-12}M_\odot$, well
within the desired range for lensing.

\begin{figure}
\plotone{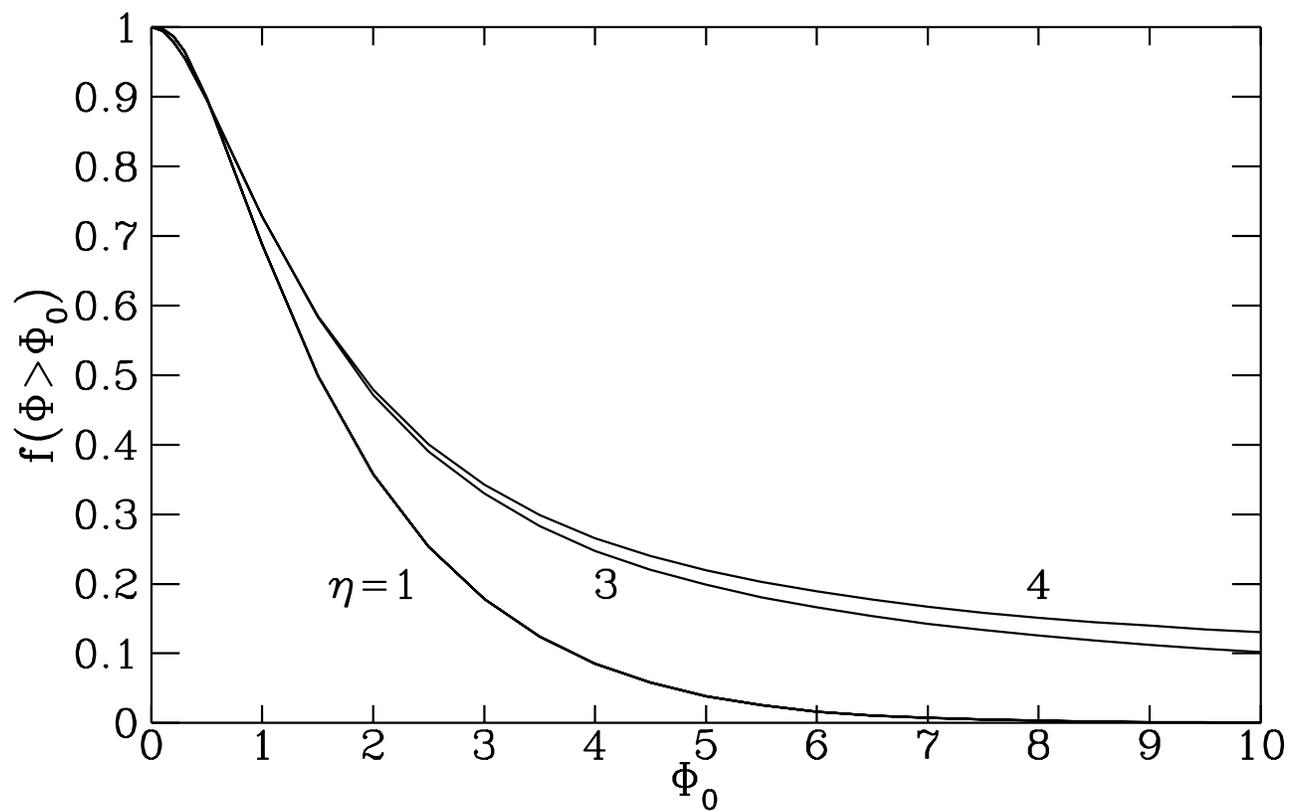}
\caption{The mass fraction of axions in miniclusters with $\Phi$
greater than $\Phi_0$ as a function of $\Phi_0$.  By $\eta=4$ the
evolution has very nearly frozen.  \label{fig2}}
\end{figure}

Using the radius of miniclusters from equation (\ref{R}) and the
Einstein ring radius from equation (\ref{re}), we find $R_E/R_{\rm mc}
\simeq 0.15\Phi \left(1+\Phi\right)^{1/3}\Omega_a h^{3/2}
\left(M/10^{-12}M_\odot\right)^{1/6}$.  The requirement that
miniclusters be smaller than their Einstein ring radius places
restrictions on $\Phi$ as a function of $M$.  We see that if
$\Omega_a\sim 1$, then miniclusters with $M \sim10^{-12} M_\odot$ and
$\Phi \ga 5 - 10$ (depending upon $h$) are within their
Einstein ring radius. Actually, objects
with the radius comparable to their Einstein radius have a better
chance to be detected as femtolenses (compared to point mass objects)
if they are in the mass range $10^{-11} M_\odot \la M \la 10^{-13}
M_\odot$ (\cite{ug95}). From Figure \ref{fig2} we see that more than
about 20\% of all dark matter axions are in miniclusters with $\Phi
\ga 5$ and will be within their ring radius.

We conclude that
if axions are the dark matter with
$\Omega_a$ close to unity, then the fraction of $\Omega$ in
miniclusters capable of femtolensing is 15\% - 20\% (see
Fig. \ref{fig2}), i. e. will be in excess of 10\%---large
enough to result in a detectable rate.

Since miniclusters in the IGM as well as in galactic halos contribute
to femtolensing events, questions such as: how are they clustered, do
miniclusters survive the epoch of galaxy formation, are they disrupted
in their encounters with stars, etc., may be
irrelevant here. Since large-$\Phi$ miniclusters are very dense, form
early, and are well separated from each other, they should escape
mutual disruption and merging.

It can be important for the present discussion that miniclusters
with $\Phi \ga 30$ relax to Bose-stars (\cite{t91}; \cite{kt93}) and
consequently became even densier and compact.

We conclude that axion miniclusters are candidates for femtolensing
and picolensing.
We do not believe that the conclusion is sensitive to specific initial
conditions, but is generic so long as there are large super-horizon-sized
fluctuations in the misalignment angle during the development of the
axion potential.

\acknowledgments

We thank S. Colombi, A. Gould and A. Stebbins for useful discussions.
This work was supported by DOE and NASA grant NAG 5-2788 at Fermilab and
DOE grant DE-AC02-76ER01545 at Ohio.

\newpage

\begin{thebibliography}{}

\bibitem[Abbott \& Sikivie 1983]{as83}
Abbott, L. F.,  \&  Sikivie,  P.  1983,
Phys.~Lett., 120B, 133

\bibitem[Battye \& Shellard 1994]{bs94}
Battye, R. A.,  \&   Shellard,  E. P. S. 1994,
\prl, 73, 2954

\bibitem[Davis 1986]{d86}
Davis,  R. 1986,
Phys.~Lett., 180B, 225

\bibitem[Deguchi \& Watson 1986]{dw86}
Deguchi, S.,  \&  Watson, W. D. 1986,
\apj, 307, 30

\bibitem[De Rujula et al 1992]{rjm92}
De Rujula, A., Jetzer,  P., \&  Mass\'{o} , E. 1992,
\aap, 254, 99

\bibitem[Dicus et al 1978]{dktw78}
Dicus, D. A., Kolb, E. W., Teplitz,  V. L., \&  Wagoner, R. V.  1978,
\prd,  18, 1829


\bibitem[Dine \& Fischler 1983]{df83}
Dine M.,  \&   Fischler, W. 1983,
Phys.~Lett., 120B, 137

\bibitem[Fukugita et al 1982]{fwy82a}
Fukugita, M., Watamura, S., \& Yoshimura, M. 1982a,
\prl,  48, 1522


\bibitem[Gould 1992]{g92}
Gould, A. 1992,
\apj,  386, L5

\bibitem[Gross et al 1981]{gpy81}
Gross, D., Pisarski, R., and Yaffe, L. 1981,
Rev.~Mod.~Phys., 53, 43

\bibitem[Harari \& Sikivi 1987]{hs87}
Harari D., \&  Sikivie, P. 1987,
Phys.~Lett.,  195, 361

\bibitem[Hatsuda \& Yoshimura 1988]{hy88}
Hatsuda, T., \& and Yoshimura, M. 1988,
Phys.~Lett.,  203B, 469

\bibitem[Hogan \& Ress 1988]{hr88}
Hogan, C. J., \& Rees, M. J.  1988,
Phys.~Lett.,  205B, 228

\bibitem[Kolb \& Tkachev 1993]{kt93}
Kolb, E. W., \& Tkachev, I. I.  1993,
\prl,  71, 3051

\bibitem[Kolb \& Tkachev 1994a]{kt94a}
Kolb, E. W., \& Tkachev, I. I. 1994a,
\prd,  49, 5040

\bibitem[Kolb \& Tkachev 1994b]{kt94b}
Kolb, E. W., \& Tkachev, I. I. 1994b,
\prd,   50, 769


\bibitem[Mandzhos 1981]{m81}
Mandzhos, A. V.  1981,
Soviet Astron.~Lett.,  7, 213

\bibitem[Mayle et al 1988]{m88}
Mayle, R., Wilson, J. R.,  Ellis, J.,
Olive, K. A., Schramm, D. N., \& Steigman, G. 1988
Phys.~Lett., 203B, 188

\bibitem[Nemiroff \& Gould 1995]{ng95}
Nemiroff, R. J., \& Gould, A. 1995
Ohio State University Preprint OSU-TA-9/95, \apj letters in press.

\bibitem[Ohanian 1983]{o83}
Ohanian, H. C.   1983,
\apj,  271, 551 (1983)

\bibitem[Peccei \& Quinn 1977]{pq77a}
Peccei, R. D., \&  Quinn, H.  1977
\prl,  38, 1440


\bibitem[Peterson \& Falk 1991]{pf91}
Peterson, J. B., \& Falk, T. 1991,
\apj,   374, L5

\bibitem[Preskill et al 1983]{pww83}
Preskill, J., Wise M., \& Wilczek, F 1983,
Phys.~Lett., 120B, 127

\bibitem[Press \& Gunn 1973]{pg73}
Press, W. H. \& Gunn, J. I. 1973,
\apj, 185, 397

\bibitem[Raffelt \& Seckel 1988]{rs88}
Raffelt, G. G., \& and  Seckel, D. 1988,
\prl,  60, 1793  (1988)

\bibitem[Raffelt 1990]{r90}
Raffelt, G. G.  1990,
\physrep, C198, 1

\bibitem[Schneider \& Schmidt-Burgk 1985]{ss85}
Schneider, P., \& Schmidt-Burgk, J. 1985,
\aap,  148, 369

\bibitem[Stanek et al 1993]{spg93}
Stanek, K. Z., Paczynski, B., \&  Goodman, J. 1993,
\apj,  413, L7

\bibitem[Tkachev 1991]{t91}
Tkachev, I. I. 1991,
Phys.~Lett.,  B261, 289


\bibitem[Turner 1986]{t86}
Turner, M. S.  1986,
\prd,  33, 889

\bibitem[Turner 1988]{t88}
Turner, M. S.  1988,
\prl,  60, 1797

\bibitem[Turner 1990]{t90}
Turner, M. S.  1990,
\physrep, C197, 67

\bibitem[Ulmer \& Goodman 1995]{ug95}
Ulmer, A., \&  Goodman, J. 1995
\apj,  442, 67

\bibitem[Weinberg 1978]{sw78}
Weinberg, S. 1978,
\prl,  40, 223

\bibitem[Wilczek 1978]{fw78}
Wilczek, F. 1978,
\prl,  40, 279




\end{thebibliography}
\end{document}